\documentclass[aps,prl,showpacs,12pt]{revtex4-1}
\usepackage{graphicx}
\usepackage{epstopdf}
\usepackage{amsmath}
\usepackage{amssymb}
\usepackage{default}
\usepackage{default}
\newcommand{\be}{\begin{equation}}
\newcommand{\bea}{\begin{eqnarray}}
\newcommand{\bc}{\begin{center}}            
\newcommand{\ee}{\end{equation}}
\newcommand{\eea}{\end{eqnarray}}
\newcommand{\ec}{\end{center}}
\newcommand{\baa}{\begin{eqnarray*}}
\newcommand{\eaa}{\end{eqnarray*}}
\begin{document}
\title{Optimal performance of heat engines with a finite source or sink \\  
and inequalities between means}
\author{Ramandeep S. Johal}
\email{rsjohal@iisermohali.ac.in}
\affiliation{Department of Physical Sciences,  
Indian Institute of Science Education and Research Mohali, 
Sector 81, S.A.S. Nagar, Manauli PO 140306, Punjab, India}
\begin{abstract}
Given a system with a finite heat capacity and a heat reservoir,
and two values of initial temperatures, $T_+$ and $T_- (< T_+)$, we enquire,  
in which case the optimal work extraction is larger:  
when the reservoir is an infinite source at $T_+$ and the system is a sink at $T_-$,
or, when the reservoir is an infinite sink at $T_-$ and the system acts as a 
source at $T_+$?
It is found that in order to compare the total extracted work, and
the corresponding efficiency in the two 
cases, we need to consider three regimes as suggested by
an inequality, the so-called arithmetic mean-geometric mean
inequality, involving the arithmetic and the 
geometric means of the two temperature values $T_+$
and $T_-$. In each of these regimes, the efficiency
at total work obeys certain universal bounds, given only in terms
of the ratio of initial temperatures.  
The general theoretical results are exemplified for thermodynamic
systems for which internal energy and temperature are power laws of
the entropy. The conclusions may serve as benchmarks in the design 
of heat engines, where we can choose the nature  of the finite system,
so as to tune the total extractable work and/or the corresponding efficiency. 
\end{abstract}
\maketitle
\section{I. Introduction}
Thermodynamics is regarded as a discipline with a formal
simplicity, but still covering a wide domain of applicability. One of the 
central problems in thermodynamics is 
the extent of heat-to-work conversion, with its focus on 
 maximal work or power output and the consequent efficiency of the process.
The seminal results of Carnot apply to the case of infinite reservoirs.
However, in recent years, the study of the role of finite reservoirs has also 
caught attention \cite{Ondrechen1981, Ondrechen1983,Leff1987, Chen1997,Izumida2014,
Wang2014, JohalRai2016}.
This is motivated by practical considerations
such as a limited supply of fuel (a finite heat source),
or the working medium being in contact with a small environment (sink)
which may be the case in small-scale devices, or even relevant for the 
design of modern cities. 

On the other hand, algebraic inequalities between 
the means hold a kind of poetic fascination. 
One of the most important \cite{Alsina} and well-known is 
the arithmetic mean-geometric mean (AM-GM) inequality, stated as follows. 
For two real positive numbers, $a$ and $b$,
with arithmetic mean $A(a,b) = (a+b)/2$
and geometric mean $G(a,b) = \sqrt{a b}$, we have 
\be
\frac{a+b}{2} \geqslant \sqrt{a b},
\label{amgm}
\ee
with equality only if $a=b$. 
Such inequalities are useful in proving  elementary
results in many disciplines \cite{Hardy52, Bullen88}. 
Especially, in the context 
of macroscopic thermodynamics, 
the second law of increase of entropy may be argued as
follows \cite{Cashwell67}.
Consider $n$ systems with a constant heat capacity
$C$ and initial temperatures, 
$\{ T_i |i=1,...,n \}$. Placed in  mutual
thermal contact, these systems come to equilibrium
at a common final temperature, say $T_f$. 
From the energy conservation condition (the first law),
we have $\sum_i C(T_i - T_f) = 0$, which implies $T_f = \sum_i T_i /n$.
Now the total entropy change:  $\Delta S =  \sum_i \int_{T_i}^{T_f} (C/T)dT
= n C (\ln T_f - \ln (\Pi_i T_i)^{1/n})$, so by virtue of the AM-GM
inequality \cite{genamgm}, we get $\Delta S \geqslant 0$ \cite{commenta, Tait1868,
Sommerfeld64, Landsberg87}.
Thus in the above argument, the manifestation of 
AM-GM inequality is specifically 
tied to the assumption of a particular model system.
By assuming systems other than
perfect gases, one can invoke inequalities between other means.

It is apparent that alternative thermodynamic processes, such 
as optimal work-extracting processes, would exhibit
a similar connection between physical models and specific 
inequalities between the means. In this paper, 
our objective is to compare the work output capacity
and efficiency of two complementary scenarios, involving
a finite system and a reservoir. During this analysis, 
we will uncover a rather general role of the AM-GM inequality. 
In particular, we will 
address the following question. Assume a 
pair of values for temperature, say $T_+$ and $T_- (<T_+)$,
and a system A with a finite heat capacity.
Also, a heat reservoir is present  
such that if the system is at temperature $T_+$,
the reservoir is a sink at $T_-$. Conversely,
if the system is at $T_-$, then the reservoir 
is a hot source at $T_+$.
Which of these two situations (see Fig. 1) would yield 
a larger amount of extractable work, due to temperature difference?
We answer this question by assuming that the process of maximal work
extraction is carried out by some working medium (whose
details are not important) via infinitesimal 
reversible heat cycles between system A and the reservoir.

In practical terms, we may consider a toy engine 
which can ideally work in a reversible manner, utilizing
the temperature gradient between system A
and the environment.  Let $T_+$ and $T_-$ be the environment
temperatures, say, in summer and in winter season, respectively. 
So in summer, we cool the system A to temperature $T_-$,
while in winter, we have to heat up the system  to temperature $T_+$,
in order to run the engine.
The engine works till it equilibrates at the specific temperature
of the environment. When will the engine yield a larger amount
of total work, in summer, or in winter?

The paper is organized as follows. In Section II, 
we describe the framework using two scenarios for work extraction 
due to temperature difference between a finite
system and a heat reservoir. In Subsection II.A,
the total extracted work and the corresponding
efficiency are compared for the two scenarios.
In Section III, physical examples are given 
based on thermodynamic systems where the temperature
and the internal energy are related to the entropy
by power laws.
Section IV discusses the bounds on the efficiency
at total work. Finally, Section V is devoted to 
summary and concluding remarks. 

\section{II. Work from a finite system and a reservoir}
To set up the thermodynamic framework, consider  system A following a certain
fundamental relation $U = U(S,V,N)$. It has equilibrium states
described by energy $U_+$, entropy $S_+$ at temperature
$T_+$, and alternatively, by $U_-$ and $S_-$ at $T_-$, with some fixed values
of volume $V$ and number of moles $N$. 
For simplicity, we consider only systems with a positive heat capacity ($C_V >0$).
This implies that $U_+ > U_-$ and $S_+ > S_-$.

Now, we first assume that system A acts as a finite heat sink at temperature $T_-$,
relative to a very large hot reservoir (source) at temperature $T_+$.
We couple the two by running infinitesimal heat cycles, which successively increase
the temperature of A, till A comes in
equilibrium with the hot source, see Fig.1 (i).
At an arbitrary intermediate stage, when the
temperature of A is $T$, the small amount of heat removed from the source $dQ_h$ 
is converted into an amount of work $dW$ with maximal (Carnot)
efficiency $\eta = 1- T/T_+$.  The heat discarded to
the sink is $dQ_c = C_V dT$. Then, we can write
$dW = \eta(1-\eta)^{-1} dQ_c$. The total extracted work
is given by:
\bea
W_+ &=& \int_{T_-}^{T_+} dW \\
 & = & \int_{T_-}^{T_+} \frac{\eta}{1-\eta} C_V dT \\
&=&  T_+ (S_+ - S_-) - (U_+ - U_-).
\label{wp}
\eea
The heat absorbed from the hot source is $Q_+ =   T_+ (S_+ - S_-)$.
Then the efficiency at total work, $\eta_+ = W_+ /Q_+$,
is calculated to be:
\be
\eta_+ = 1- \frac{1}{T_+} \frac{U_+ - U_-}{S_+ - S_-}.
\label{ep}
\ee
Then, we consider the alternative situation in which  
A acts as a finite source at temperature $T_+$,
relative to an infinite sink at $T_-$, see Fig.1 (ii). Again, we
extract the maximal work by utilizing the temperature 
gradient between A and the reservoir, till A  
is at temperature $T_-$. Then,
after a similar calculation \cite{Izumida2014} as above, 
the total work obtained is
\be
W_- = (U_+ - U_-) -  T_- (S_+ - S_-).
\label{wm}
\ee
This is termed as exergy in the engineering literature \cite{Exergy}.
The heat absorbed from the source is $Q_- = U_+ - U_-$,
while the efficiency of the process $\eta_- = W_- /Q_-$ is given by
\be
\eta_- = 1- T_-\frac{ S_+ - S_-}{U_+ - U_-}.
\label{em}
\ee
\begin{figure}
  \includegraphics[width=13cm]{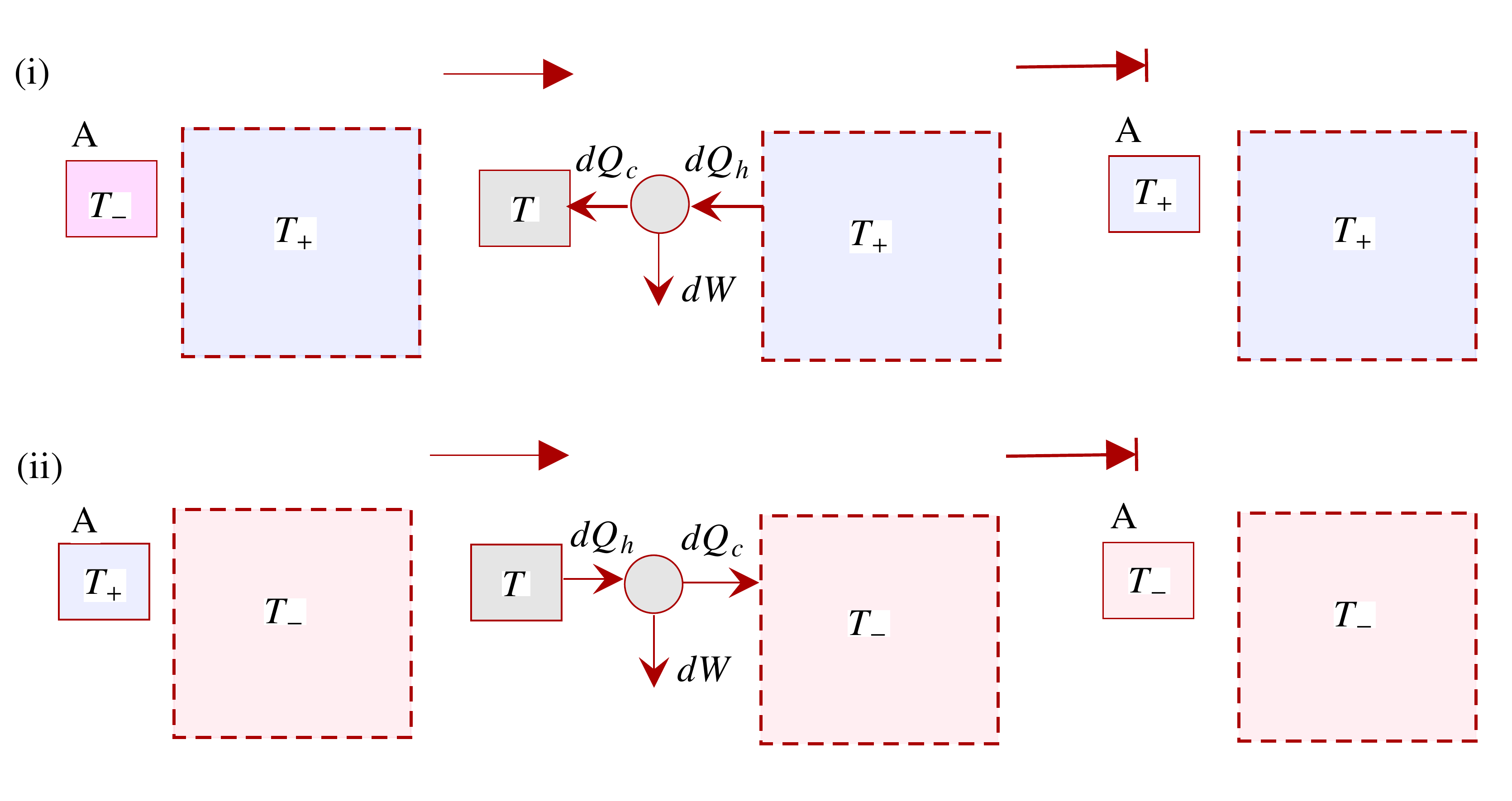}
  \caption{Schematic of the reversible heat engine between a finite system and a heat reservoir, 
  for a given pair of initial temperatures $(T_+,T_-)$: 
  (i) System A is a finite sink at $T_-$ and is coupled to an infinite source at $T_+$,
  via heat engine. Work extraction $W_+$, Eq. (\ref{wp}),
  is completed when the temperature
  of A becomes $T_+$. (ii) System A is a finite source at $T_+$ and 
  is coupled to an infinite sink
  at $T_-$, via heat engine. Total 
  extracted work is $W_-$, Eq. (\ref{wm}), when the temperature of A becomes $T_-$.}
 \end{figure}
Thus for the toy engine mentioned in Introduction, 
$W_+$ and $\eta_+$ ($W_-$ and  $\eta_-$) may  
refer to the total work and the corresponding
efficiency in summer (winter) season.

\subsection{A. The Comparison}
Now we compare the amounts of extracted work, 
and the efficiencies, in these alternative set-ups.
For that purpose, we recall the classic
result in calculus, known as the {\it mean value theorem}. Consider a
continuous and differentiable function $U(S)$ in the domain
$[S_-,S_+]$, with the derivative $T(S) = dU/dS$. 
Let us denote:  $U(S_{\pm}) = U_{\pm}$. 
Following the theorem, there is a point 
$S_m$ strictly within this interval ($S_+ > S_m > S_-$), 
at which the derivative of the function $U$,
i.e. $T(S_m) \equiv T_m$, is given by:
\be
 T_m = \frac{U_+ - U_-}{S_+ - S_-}.
\label{tm}
 \ee
We also assume $T(S)$ to be monotonic
increasing function, or, in other words, $U(S)$ is 
a convex function. In the context of thermodynamics,
this assumption implies
positive heat capacity ($C_V$) of the system. 
Then it follows that $ T(S_+) > T(S_m) > T(S_-)$, or alternatively,
$ T_+ > T_m > T_-$. 

Now, depending 
on the nature of the thermodynamic system i.e. the form of the function $U(S)$, 
$T_m$ can take values relative to $A(T_+, T_-)$ and $G(T_+, T_-)$, 
such that we have the following situations:
\bea
(a) \qquad\qquad  T_+ &>&  T_m \geqslant  \frac{T_+ + T_-}{2} >
\sqrt{T_+ T_-} > T_- \nonumber \\
(b) \qquad\qquad  T_+ &>&   \frac{T_+ + T_-}{2} > T_m > 
\sqrt{T_+ T_-} > T_- \nonumber  \\ 
(c) \qquad\qquad T_+ &>&   \frac{T_+ + T_-}{2} > \sqrt{T_+ T_-}
\geqslant T_m   > T_-  \nonumber \\
\label{abc}
\eea
We choose the means $A$ and $G$ to split the interval $(T_-,T_+)$ into 
three regions, because for $T_m = (T_+ + T_-)/2$, we have 
$W_+ = W_-$, and for $T_m = \sqrt{T_+ T_-}$, we have $\eta_+ = \eta_-$.
This helps naturally to compare the magnitudes of work, and efficiency.
Thus, if case $(a)$ holds, then applying $T_m \geqslant (T_+ + T_-)/2$,
and using Eqs. (\ref{tm}), (\ref{wp}) and (\ref{wm}), we obtain
$W_+ \leqslant W_-$. In this case, due to AM-GM inequality, 
we also have  $T_m > \sqrt{T_+ T_-}$, 
which implies  $\eta_+ < \eta_-$,  due to Eqs. (\ref{tm}), 
(\ref{ep}) and (\ref{em}).   

Similarly, if case $(b)$ applies, then we conclude 
that  $W_+ > W_-$, but due to AM-GM inequality, we  have 
$\eta_+ < \eta_-$.
If case $(c)$ is true, i.e.  $\sqrt{T_+ T_-} \geqslant T_m$,
it implies $\eta_+ \geqslant \eta_-$. Further, due to 
 $(T_+ + T_-)/2 >  T_m$, we also have
$W_+ > W_-$. The above three scenarios are summarized in Table I.

Thus we see that the comparison of $T_m$ with $A(T_+,T_-)$
decides the relative magnitudes of $W_+$ and $W_-$, whereas
the comparison of $T_m$ with $G(T_+,T_-)$, 
serves to compare $\eta_+$ and $\eta_-$.
In these comparisons, the AM-GM inequality provides a 
sort of background against which $T_m$ takes
values depending on the nature of system A (see examples below).
In terms of practical utility, the goal behind
modelling of heat engines is to characterize their 
optimal working regimes. In this regard, if we are given 
a finite system A and a constraint to run the engine 
in one of the two scenarios, denoted as (i) and (ii) in the above,
then a particular choice can be motivated as follows.
In case the system A falls in category (a) of Table I, then
choice (ii) provides a higher total work output and a higher
efficiency. On the other hand, if system A belongs to 
category (c), then the choice (i)  
would provide a higher work output and a higher 
efficiency. In case the system belongs to regime (b), we have
a situation with a trade-off. If we opt for a higher work output then 
the efficiency obtained is less, and vice versa.  
Heuristically, one may be able to make a choice in 
this situation as follows.
A focus on a higher efficiency may become important, if
the substance (system A) is in short supply or if the
economic/ecological costs of preparing the system, in the desired
state, are rather high. On the other hand, if such 
costs are not a consideration, then one may  
focus on higher total work, with the corresponding 
efficiency being less of a concern.

\begin{table}
\begin{tabular}{|c|c|c|}
\hline
$(a)$   &   $(b)$    &   $(c)$   \\
\hline
\quad $W_+ \leqslant W_-$  \quad     &  \quad $W_+ > W_-$  \quad 
& \quad $W_+ > W_-$  \quad  \\
\quad $\eta_+ < \eta_-$ \quad &  \quad  $\eta_+ < \eta_-$ \quad    &
\quad $\eta_+ \geqslant \eta_-$ \quad\\
\hline
\end{tabular}
\caption{Comparison of total work, Eqs. (\ref{wp}) and (\ref{wm}), 
and efficiency at total work,  Eqs. (\ref{ep}) and (\ref{em}), 
corresponding to regimes $(a), (b)$ and $(c)$ in Eq. (\ref{abc}).}
\end{table}

\section{III. Examples}
In this section,
we illustrate the various cases noted in the above, by taking examples from 
different types of physical systems.
Consider a class of thermodynamic systems 
that obey: $U \propto S^{\omega}$ and  $T \propto S^{\omega -1}$, where
$\omega$ is a constant real number. For heat capacity
to be positive, we must have $\omega >1$.
So, $T_m$ is evaluated to be:
\be
T_m =  \frac{1}{\omega}
\frac{T_{+}^{\omega/(\omega -1)} - T_{-}^{\omega/(\omega -1)}}
{T_{+}^{1/(\omega -1)}-T_{-}^{1/(\omega -1)}}.
\label{ttw}
\ee
It is convenient to introduce the 
generalized mean \cite{Stolarsky75,Alzer87} of two real, 
positive numbers $(a,b)$:
\be
E_r(a,b)  = \frac{r-1}{r} \frac{a^r-b^r}{a^{r-1}-b^{r-1}}.
\ee 
In our case, $T_m = E_r(T_+,T_-)$  with $r = \omega/(\omega -1)$.
For $r=2 \; (\omega=2)$, 
 $E_2( T_+,T_-) = (T_+ + T_-)/2$. For $r=1/ 2$ $(\omega=-1)$, 
 $E_{1/2}(T_+,T_-) = \sqrt{T_+,T_-}$.
Since $E_r(a,b)$ is increasing in parameter $r$ \cite{YangCao},
 it follows that, for $r\geqslant2$ or $\omega \geqslant 2$, 
 we have $T_m = E_r(T_+,T_-) \geqslant E_2(T_+,T_-)$, 
 which  implies $T_m \geqslant (T_+ + T_-)/2$, or case $(a)$.  
Therefore, for $2  > \omega > 1$, the system corresponds to case $(b)$.   

Some examples of physical systems in the above class, for appropriate
values of $T_+$ and $T_-$, are:
$\omega = 4/3$ (black-body radiation), $\omega =5/3$ (degenerate Bose gas)
and $\omega = 2$ (ideal Fermi gas).
The case of a perfect-gas system, can be discussed as 
the limit $r\to 1$, which yields $E_1(T_+,T_-) = L(T_+,T_-)$, known as
 the logarithmic mean \cite{Carlson72,Bhatia08}:
 \be
L(T_+,T_-) =\frac{T_+ - T_-}{\ln T_+ - \ln T_-}.
\ee
Logarithmic mean temperature difference is a useful measure
of the effectiveness with which a heat exchanger can transfer
heat energy \cite{Nedderman1985}. This mean satisfies:
\be
\frac{T_+ + T_-}{2}  > L(T_+,T_-) > \sqrt{T_+ T_-}.
\label{lmin}
\ee
So if $T_m = L(T_+,T_-)$, then due to the above inequality, we have 
an instance of case $(b)$. Thus with a perfect-gas system,  the 
finite-sink/infinite-source setup produces
more work than finite-source/infinite-sink setup ($W_+ > W_-$),
although the efficiency at total work follows the reverse order 
($\eta_+ < \eta_-$).

As our final model system,
let A consist of $N$ non-interacting, localized spin-1/2 particles 
\cite{Pathria}.
Each particle can be regarded as a two-level system,  
with energy levels ($0$, $\epsilon$).   
The mean energy for this system, in the limit of high temperatures such that
$\epsilon \ll k T$, on keeping terms only upto $(\epsilon/k T)^2$, 
can be approximated as:
$U \approx N ({\epsilon}/{2}-{\epsilon^2}/{4kT})$, with entropy 
$S \approx N k (\ln 2 - {\epsilon^2}/{8k^2T^2})$. Then from Eq. (\ref{tm}),
we have: $T_m = 2T_+ T_-/(T_+ + T_-)$, which is the well-known
harmonic mean $H(T_+,T_-)$. This mean is strictly less than 
$G(T_+,T_-)$, and thus our spins-system lies in regime $(c)$. 

\section{IV. Bounds on efficiency}
So far, we have noted the comparison between work 
characteristics for the two given scenarios. In
the following, we point out that within  
a given scenario, the efficiency at total extracted work 
obeys definite bounds, which are specific to 
each of the regimes $(a), (b)$ and $(c)$.
Thus if $T_m \geqslant (T_+ + T_-)/2$,
then we get from Eq. (\ref{ep}), $\eta_+ \leqslant \eta_C /2$
where $\eta_C = 1- T_-/T_+$ is the Carnot limit.
Also from Eq. (\ref{ep}), we get $\eta_- \geqslant \eta_C /(2-\eta_C)$.
Similarly, in regime (c), when $T_m \leqslant \sqrt{T_+ T_-}$,
we get $\eta_{+} \geqslant \eta_{CA}$ and $\eta_{-} \leqslant \eta_{CA}$, 
where $\eta_{CA} = 1 - \sqrt{T_-/T_+}$ \cite{Chambadal, Novikov},
which is popularly known as CA-efficiency, after F. L. Curzon and 
B. Ahlborn who rediscovered this formula \cite{CA1975},
see also \cite{Feidt2014}.
These comparative bounds are summarized in Table II, as well
as they are depicted in Figs. 2 and 3. Note that  
the efficiencies $\eta_C /2$, $\eta_{CA}$ and $\eta_C /(2-\eta_C)$
are frequently discussed in the context of maximum power output 
in finite-time models 
\cite{Broeck2005, Chambadal, Novikov, CA1975, Esposito2010}.
But we observe that, here, within a quasi-static framework, $\eta_{CA}$
serves to separate $\eta_+$ and $\eta_-$ in regimes $(b)$ and $(c)$.

\begin{table}
\begin{tabular}{|c|c|c|c|}
\hline
$(a)$   &   $(b)$    &   $(c)$   \\
\hline
   \quad $ 0< \eta_+ \leqslant \eta_C /2 $  \quad     & 
 \quad $\eta_C /2 < \eta_+ < \eta_{CA}$  \quad  & \quad $ \eta_{CA} 
 \leqslant \eta_+ < \eta_C $  \quad  \\
 \hline 
 \quad $ \frac{\eta_C}{2-\eta_C} \leqslant \eta_- < \eta_C $  \quad     & 
 \quad $ \eta_{CA} < \eta_- < \frac{\eta_C}{2-\eta_C} $  \quad   
 & \quad  $ 0 <  \eta_- \leqslant \eta_{CA}   $  \quad  \\
\hline
\end{tabular}
\caption{The bounds obeyed by efficiencies at total extracted work, 
$\eta_+$ and $\eta_-$,
in respective regimes given in Eq. (\ref{abc}), where 
 $\eta_C = 1 -T_-/T_+ $ and 
 $\eta_{CA} = 1 -\sqrt{T_-/T_+}$.} 
\end{table}

\begin{figure}[ht]
  \includegraphics[width=9cm]{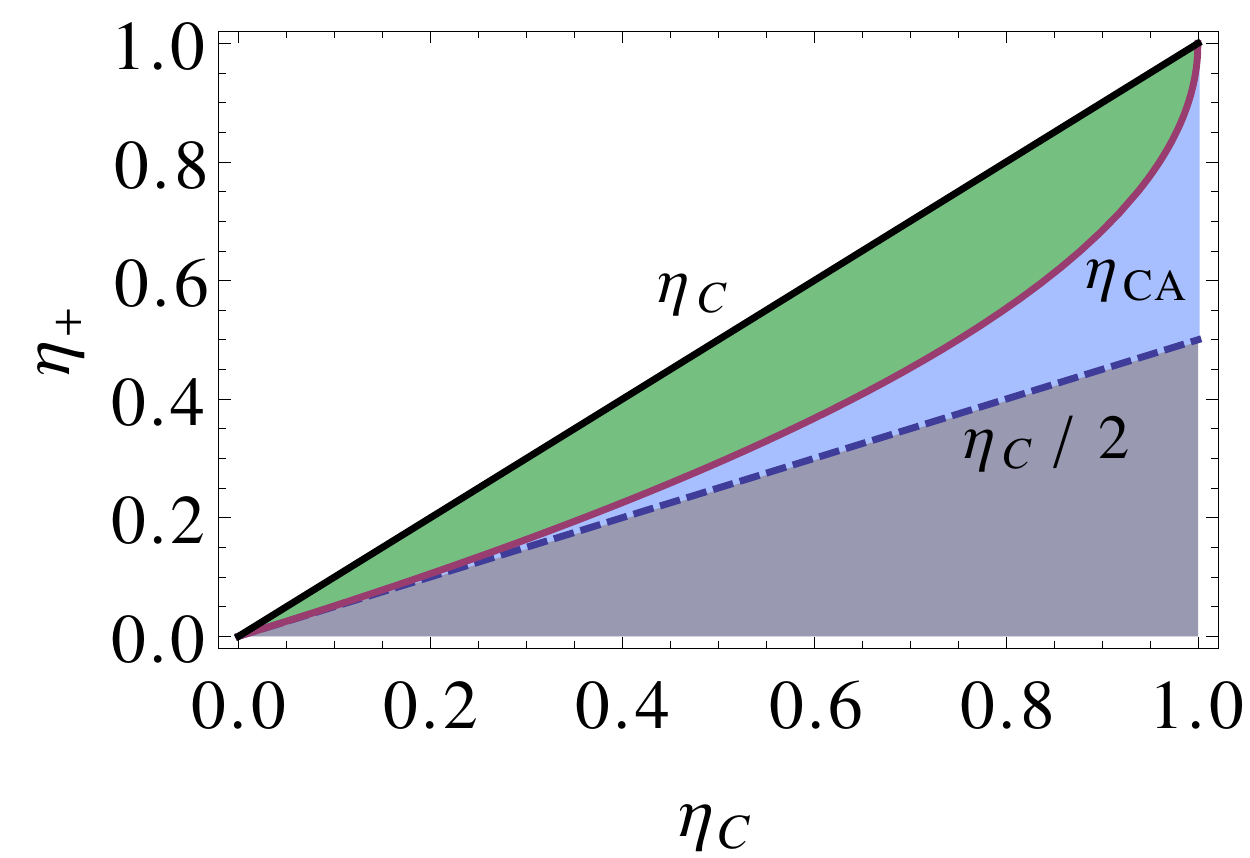} 
  \caption{Bounds on efficiency $\eta_+$, in  
  the regimes (from bottom to top) $(a), (b)$, and $(c)$,
  as given in Table II.}
 \end{figure}

 \begin{figure}[ht]
  \includegraphics[width=9cm]{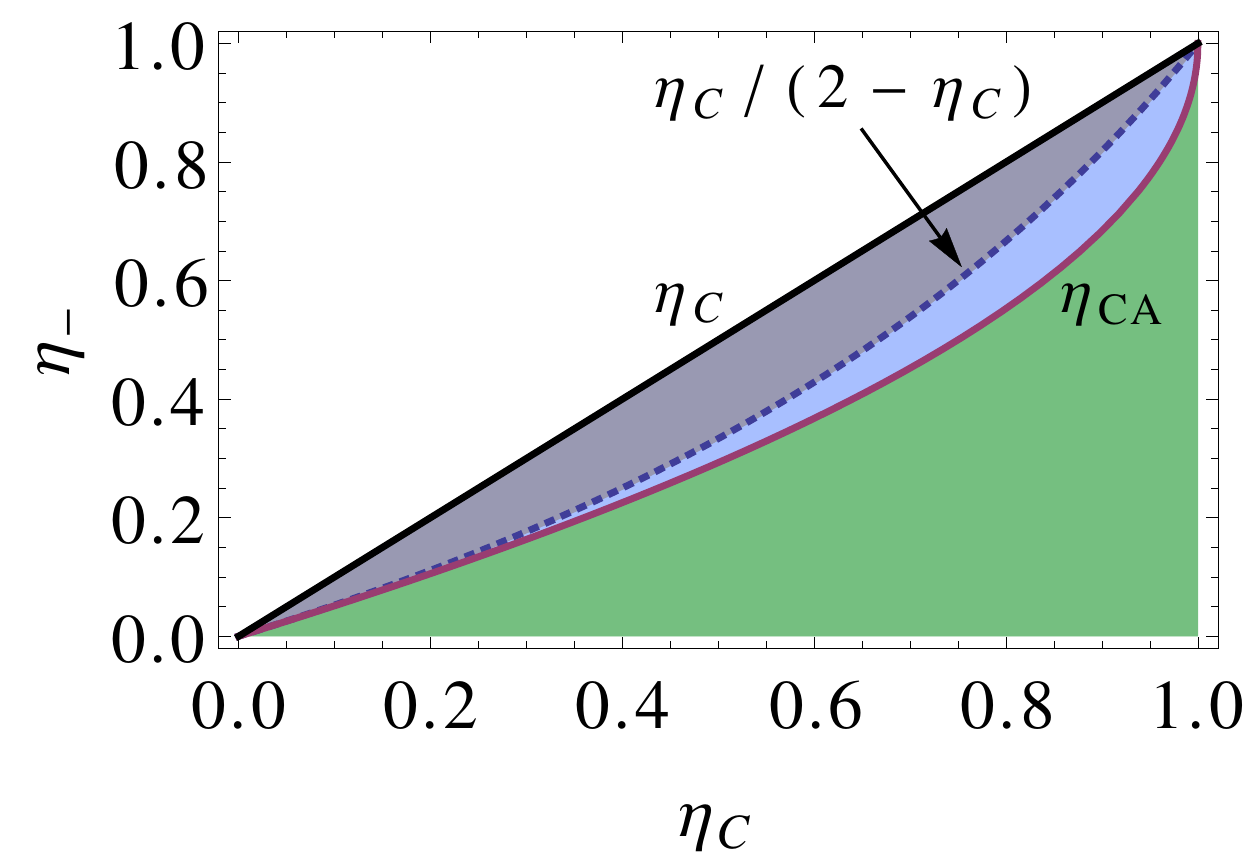} 
  \caption{Bounds on efficiency $\eta_-$, in  
  the regimes (from top to bottom) $(a), (b)$, and $(c)$, as in Table II.}
 \end{figure}
The above bounds are universal as they depend only 
on the ratio of the initial temperatures.
Note that the actual expressions, (\ref{ep}) and (\ref{em}),
do depend, in general, on 
the nature of system A. But 
close to equilibrium, even the general expressions for $\eta_+$
and $\eta_-$ exhibit a universality.
Thus assuming linear response, we can expand
energy upto second order in the entropy difference $\delta S =
S_+ - S_-$ \cite{JohalRai2016}:
\be 
U(S_-) = U(S_+) -T_+ \delta S
             + \frac{1}{2} \left. \frac{d T}{d S} \right|_{S= S_+}
             \hspace{-5mm}(\delta S)^2. 
\label{usm2}
\ee
Using the above expansion in Eq. (\ref{tm}), and upon simplifying, we get
$T_m = (T_+ + T_-)/2$. This implies
that $W_+ = (T_+ - T_-)\delta S /2 = W_-$. Thus, under linear response,
the extracted work is same in both the cases. 
However, the efficiency at total work is approximated as  
$\eta_+  =  \eta_C /2$ and $\eta_- = {\eta_C}/(2-\eta_C)$.
These expressions are consistent with the findings of Ref.
\cite{JohalRai2016}, where the lower and the upper 
bounds for efficiency with unequal-sized source and sink,
obey the same expressions. 

\section{V. Concluding remarks}
We close this investigation by making a few remarks.
Apart from an entropy-conserving process, we may analyze an 
energy-conserving process. 
The initial and final situations are the same as (i) and (ii) in Fig.1.
Specifically, for situation (i), an amount of heat energy $U_+-U_-$ is removed
quasi-statically from the reservoir and deposited in the same manner 
with the cold system. The change in entropy of system A is $(S_+ - S_-) > 0$.
The change in entropy of the reservoir is:  $-(U_+ - U_-)/T_+$. 
Thus the total change in the entropy of the universe is:
\be
\Delta S_+ = (S_+ - S_-) - \frac{U_+ - U_-}{T_+}.
\label{dsp}
\ee
Similarly, if we consider situation (ii), we can conclude 
that the total entropy change of the universe, in an energy-conserving process, 
would be:
\be
\Delta S_- = -(S_+ - S_-) + \frac{U_+ - U_-}{T_-}.
\label{dsm}
\ee
Now, if we wish to compare the entropy production in the above two cases,
then we are led to consider the following situations:
\bea
(a') \qquad\qquad  T_+ &>&  T_m \geqslant  \frac{2 T_+ T_-}{T_+ + T_-} 
> T_- \nonumber \\
(b') \qquad\qquad  T_+ &>&   \frac{2 T_+ T_-}{T_+ + T_-}> T_m > T_-.  
\eea
It is easy to see that if case $(a')$ is true, then $\Delta S_- 
\geqslant \Delta S_+$.
The inverse inequality is valid, if case $(b')$ holds.  Thus for 
an energy-conserving process, we see that the inequality between
generalized mean $T_m$, and $H(T_+,T_-)$, quantifies 
the relative magnitudes of $\Delta S_-$ and $\Delta S_+$.

Finally, we consider an interesting meaning  of $T_m$, 
given by Eq. (\ref{tm}), in the sense of an effective temperature.
Take two heat reservoirs with temperatures
$T_m$ and $T_- (<T_m) $. Let $Q_m =  U_+ - U_-$,
be the heat extracted by the working medium from the hot reservoir in a reversible
cycle. Here $U_{\pm}$ refer to the energies of the working medium.
Then the change in entropy of the hot reservoir is $T_m Q_m = S_+ - S_-$.
The total extractable work in a reversible cycle is then 
 $(T_m - T_-)(S_+ -S_-)$, which is the same 
as $W_+$ in Eq. (\ref{wm}). The Carnot efficiency of this process
is $\eta_m = 1-T_-/T_m$, which is  Eq. (\ref{em}).
A similar conclusion follows for the other scenario,
when we consider two heat reservoirs at temperatures 
$T_+$ and $T_m (< T_+)$.
Thus $T_m$ serves as the effective temperature of one of
the two heat reservoirs in an equivalent reversible cycle,
which extracts the same amount of work and with the same
(Carnot) efficiency. 

Concluding, the main focus of this paper was the comparison
of performance of a reversible heat engine operating between
a finite system and an infinite reservoir, by switching 
 the role of the source and the sink. We compared the total
extracted work in the two cases, and the corresponding 
efficiency of the engine at those values of the work.
Interestingly, we find that the conditions for  
comparison are determined by basic mathematical inequalities
between the means, in particular the AM-GM inequality. 
The present instance of this inequality  
does not depend specifically on the nature of the system as
was the case in earlier studies. The efficiency at total work 
is naturally split into 
three regimes, based on this inequality.
The bounds separating these regimes are 
variously given as $\eta_C /2$, $\eta_{CA}$ and $\eta_C /(2-\eta_C)$.
This highlights a new significance of these expressions for efficiency,
which are usually discussed in regard to power output optimization in
finite-time models.
The utility of our conclusions may also be discussed
in the context of the toy engine mentioned in the 
Introduction. Thus, for a given pair of temperatures $(T_+, T_-)$,
we can characterize system A, or our device, based on
the regime $(a), (b)$ or $(c)$, to which it corresponds.
This determines how $W_+$ and $W_-$ compare with each other,
which further guides whether $\eta_+$ will be greater,  
or lesser, relative to $\eta_-$. Moreover, in a particular
regime, we know from Table II, the bounds within which the efficiency 
at total work is located. Thus given a choice
of system A, the efficiency at total work is restricted within a certain
range. Although derived for quasi-static processes,
these bounds may serve as benchmarks for tuning the 
performance of real devices, and can be a useful element 
in their design.

One of the limitations of our analysis may be that
we have considered idealized quasi-static processes.
In practical cases, the engines and other thermodynamic
machines work in finite cycle-times. Thus an 
extension of our analysis within an irreversible
framework  \cite{Izumida2014} may help to see how the above conclusions   
are retained or modified in finite-time models, 
at least under linear response or beyond that \cite{JohalRai2016}. 
Another interesting line of enquiry seems to be 
the connection of the bounds on efficiency
with the principles of inductive inference \cite{JRM2015,George2015}.
Finally, it is hard to ignore the aesthetic motivation
in revealing other inequalities, possibly new,
with these investigations. But, this is left 
for future work.

\section{Acknowledgements}
The author wishes to thank Dr. Renuka Rai, 
for discussions, and Jannat, for sparing her 
blackboard.


\begin{thebibliography}{32}
\bibitem [{gen()}]{genamgm}%
  \BibitemOpen
  \href@noop {} {\bibinfo  {journal} {For $n$ real positive numbers $\{ T_i
  |i=1,...,n \}$, AM-GM inequality is: $\sum_i T_i /n \geqslant (\Pi_i
  T_i)^{1/n}$}\ }\BibitemShut {NoStop}%
\end{thebibliography}
\end{document}